\documentclass[prd,aps,showpacs,superscriptaddress,floatfix,twocolumn]{revtex4-1}
\usepackage[english]{babel}
\usepackage[utf8]{inputenc}
\usepackage{float}
\usepackage{amsmath,amssymb,graphicx,textcomp}
\usepackage{hyperref}

\newcommand{\tmop}[1]{\ensuremath{\operatorname{#1}}}

\begin{document}

\date{October 10, 2016}

\title{Connected and leading disconnected hadronic light-by-light contribution
to the muon anomalous magnetic moment with physical pion mass}

\newcommand{\RBRC}{
  RIKEN BNL Research Center,
  Brookhaven National Laboratory,
  Upton, New York 11973,
  USA}

\newcommand{\UCONN}{
  Physics Department,
  University of Connecticut,
  Storrs, Connecticut 06269-3046,
  USA}

\newcommand{\NAGOYA}{
  Department of Physics,
  Nagoya University,
  Nagoya 464-8602,
  Japan}
  
\newcommand{\NISHINA}{
  Nishina Center,
  RIKEN,
  Wako, Saitama 351-0198,
  Japan}

\newcommand{\BNL}{
  Physics Department,
  Brookhaven National Laboratory,
  Upton, New York 11973,
  USA}

\newcommand{\CU}{
  Physics Department,
  Columbia University,
  New York, New York 10027,
  USA}

\newcommand{\KEK}{
  KEK Theory Center,
  Institute of Particle and Nuclear Studies, 
  High Energy Accelerator Research Organization (KEK),
  Tsukuba 305-0801,
  Japan
}

\author{Thomas Blum}
\affiliation{\UCONN}
\affiliation{\RBRC}

\author{Norman Christ}
\affiliation{\CU}

\author{Masashi Hayakawa}
\affiliation{\NAGOYA}
\affiliation{\NISHINA}

\author{Taku Izubuchi}
\affiliation{\BNL}
\affiliation{\RBRC}

\author{Luchang Jin}
\affiliation{\CU}

\author{Chulwoo Jung}
\affiliation{\BNL}

\author{Christoph Lehner}
\affiliation{\BNL}

\begin{abstract}
  We report a lattice QCD calculation of the hadronic light-by-light contribution to the muon anomalous magnetic moment at physical pion mass.  The calculation includes the connected diagrams and the leading, quark-line-disconnected diagrams. We incorporate algorithmic improvements developed in our previous work. The calculation was performed on the   $48^3 \times 96$ ensemble generated with a physical-pion-mass and a 5.5 fm spatial extent by the RBC and UKQCD collaborations using the chiral, domain wall fermion (DWF) formulation.  We find $a_\mu^{\text{HLbL}} = 5.35 (1.35)  \times 10^{- 10}$, where the error is statistical only. The finite-volume and finite lattice-spacing errors could be quite large and are the subject of on-going research. The omitted disconnected graphs, while expected to give a correction of order 10\%, also need to be computed.
\end{abstract}

\maketitle

\section{Introduction}

The lattice calculation of the hadronic light-by-light contribution to the muon anomalous magnetic moment is part of the on-going effort to better understand the approximately three standard deviation difference between the extremely accurate BNL E821 experimental result and the current theoretical calculation with similar accuracy~\cite{Jegerlehner:2009ry}. The muon anomalous magnetic moment is characterized by the small dimensionless number $a_{{\mu}} = (g_{{\mu}} - 2) / 2$, the muon anomaly.   Here the $g$-factor $g_{{\mu}}$ determines the magnetic moment of muon, $\vec{{\mu}} = \vec{s} g_{{\mu}} e / 2 m_{{\mu}}$ where $\vec{s}$ is the spin angular momentum of the muon.  The muon anomaly can be determined from the form factor $F_2$ which appears in the matrix element of the electromagnetic current:
\begin{eqnarray}
  && \bigl\langle {\mu} (\vec{p}\,') | J_{\nu} (0) | {\mu} (\vec{p}) \bigr\rangle \\
  && \hskip .5cm = - e
  \bar{u} (\vec{p}\,') \left( F_1 (q^2) \gamma_{\nu} + i \frac{F_2 (q^2)}{4 m}
  [\gamma_{\nu}, \gamma_{\rho}] q_{\rho} \right) u (\vec{p}), \nonumber
\end{eqnarray}
where $a_{{\mu}} = F_2 (0)$. Here $J_{\nu} (0)$ is the electromagnetic current, $| {\mu} (\vec{p}) \rangle$ and $| {\mu} (\vec{p}\,') \rangle$ the initial and final muon states, $q = p' - p$, and Euclidean-space conventions are used.  

A particle's anomalous magnetic moment results from its extended spatial structure.  For an elementary Dirac particle, such as an electron, muon or tau lepton, with only electroweak interactions, such structure will arise from the electroweak interactions themselves.  These effects can be computed with high precision using perturbation theory, with the leading term being the well known result of Schwinger: $a=\alpha/{2\pi}$~\cite{Schwinger:1948iu} where $\alpha$ is the fine structure constant.  However, new, high-energy phenomena that appear at an energy scale $\Lambda$ can introduce additional structure, leading to new contributions to $a_l$ that are suppressed by the ratio $(m_l/\Lambda)^2$ where $l=e$, $\mu$ or $\tau$ and $m_l$ is the mass of the corresponding lepton.  The muon anomaly may be the best place to search for such phenomena since $a_\mu$ can be more accurately measured than $a_\tau$ while $m_\mu$ is 207 times larger than $m_e$. 

The current result of the BNL experiment E821 is $a_{{\mu}}^{\exp} = 11659208.0 (6.3)
\times 10^{-10}$ \cite{Bennett:2006fi}.  More accurate experiments are planned at Fermilab (E989) \cite{Carey:2009zzb} and J-PARC (E34) \cite{Mibe:2010zz}, which aim to reduce the
error by a factor of four. Theoretically, the contributions to $g_\mu-2$ can be divided into four categories.  The first is the QED contribution, which is the largest~\cite{Aoyama:2012wk}.   The second is the electroweak correction, which is small but not negligible~\cite{Gnendiger:2013pva}.   Both the QED and electroweak contribution can be computed with perturbation theory and the uncertainties are very small. 

\begin{figure}[!htbp]
  \resizebox{0.45\columnwidth}{!}{\includegraphics{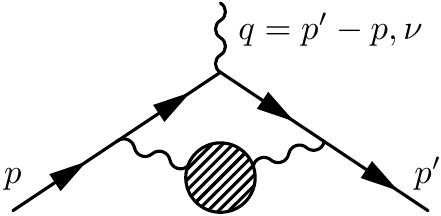}} \
  \resizebox{0.45\columnwidth}{!}{\includegraphics{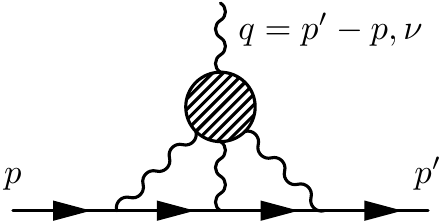}}
  \caption{\label{fig:hvp-hlbl}Left: the hadronic vacuum polarization (HVP)
  diagram. Right: the hadronic light-by-light (HLbL) diagram.  The muon and photon lines 
  are shown explicitly while the quark loops and exchanged gluons of QCD are 
  represented by the shaded circles.}
\end{figure}

The third and fourth contributions enter at second and third order in $\alpha$ and involve virtual quark loops, introducing the non-perturbative challenges of QCD.  The third is the hadronic vacuum polarization (HVP) contribution which enters at order $\alpha^2$ and corresponds to the left diagram in Fig.~\ref{fig:hvp-hlbl}. The fourth is the hadronic light by light (HLbL) contribution, corresponds to the right diagram in Fig.~\ref{fig:hvp-hlbl} and enters at order $\alpha^3$.  

The HVP contribution is the largest hadronic contribution and can be computed from a dispersion relation and experimental $e^+ e^-$ annihilation data.  This is a well-developed method with a fractional-percent error. The leading-order HVP contribution is  $692.3 (4.2) \times 10^{-10}$ \cite{Davier:2010nc} or $694.9 (4.3) \times 10^{- 10}$ \cite{Hagiwara:2011af}.  This dispersive approach is an active research area and results with reduced errors should be expected~\cite{Grange:2015fou}.  The HVP contribution can also be calculated with lattice QCD.  With recently developed methods and increased computational power, similar or even higher precision results may be possible
\cite{Bernecker:2011gh, Feng:2013xsa, Blum:2015you, Blum:2016xpd, Chakraborty:2016mwy}. 
In contrast, the HLbL contribution is at present only estimated by model calculations which give a result of $10.5 (2.6) \times 10^{- 10}$ \cite{Prades:2009tw,Benayoun:2014tra} or $11.6(3.9) \times 10^{-10}$~\cite{Jegerlehner:2009ry}. This method is difficult to improve further although it is possible to compare the model result for hadronic light-by-light scattering with a lattice result for this scattering amplitude~\cite{Green:2015sra}.  A dispersion relation analysis of the HLbL contribution is not available although work is underway in this direction~\cite{Colangelo:2014dfa,Colangelo:2014pva,Pauk:2014jza,Pauk:2014rfa,Colangelo:2015ama,Nyffeler:2016gnb}.

Combining these results gives the standard model prediction $a_\mu^{\mathrm{sm}} = 11659184.0(5.9)\times 10^{-10}$  which differs from the experimental value above by $a_\mu^{\mathrm{exp}} - a_\mu^{\mathrm{sm}} =24.0(6.9)\times 10^{-10}$, about twice the estimate for the HLbL contribution.  Thus, a systematically improvable, lattice determination of the HLbL contribution is needed to resolve or firmly establish the discrepancy. 

\begin{figure}[!htbp]
  \resizebox{0.45\columnwidth}{!}{\includegraphics{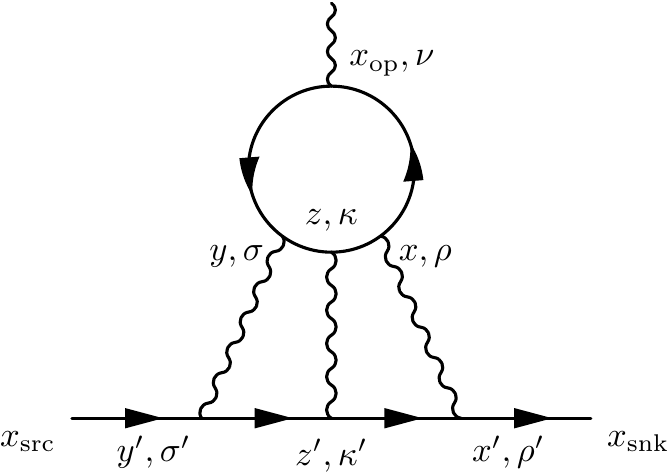}}
  \resizebox{0.45\columnwidth}{!}{\includegraphics{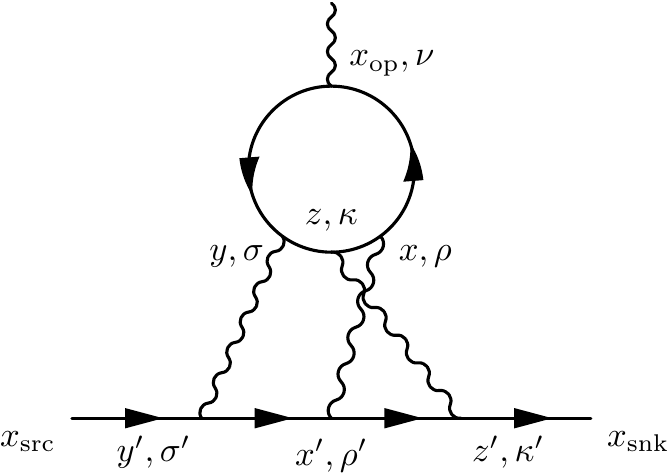}}
  \caption{\label{fig:clbl}Connected hadronic light-by-light diagrams. There
  are four additional diagrams resulting from further permutations of the
  photon vertices on the muon line.}
\end{figure}

\begin{figure}[!htbp]
  \resizebox{0.5\columnwidth}{!}{\includegraphics{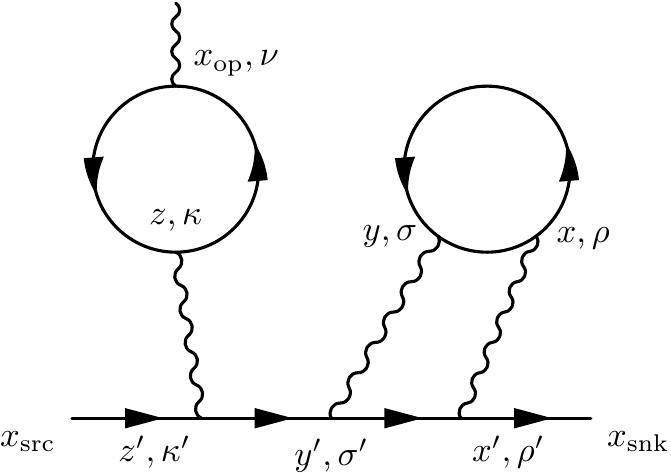}}
  \caption{\label{fig:dlbl-0}Leading-order disconnected diagram which is non-zero in
  SU(3) limit. There are additional diagrams which can be obtained from permutation of the photon vertices on the muon line.}
\end{figure}

\begin{figure}[!htbp]
  \resizebox{0.40\columnwidth}{!}{\includegraphics{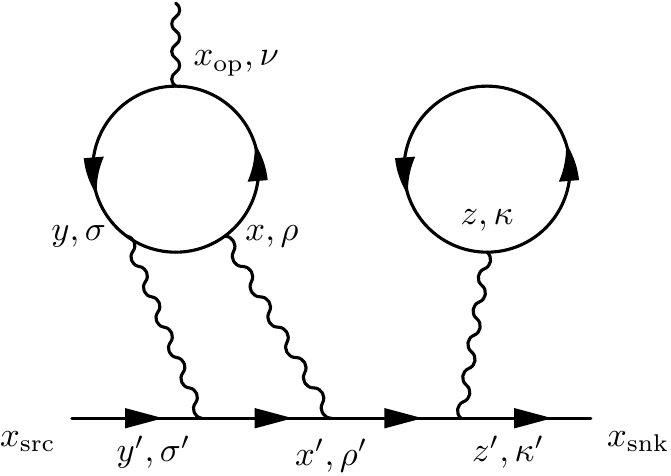}} \
  \resizebox{0.40\columnwidth}{!}{\includegraphics{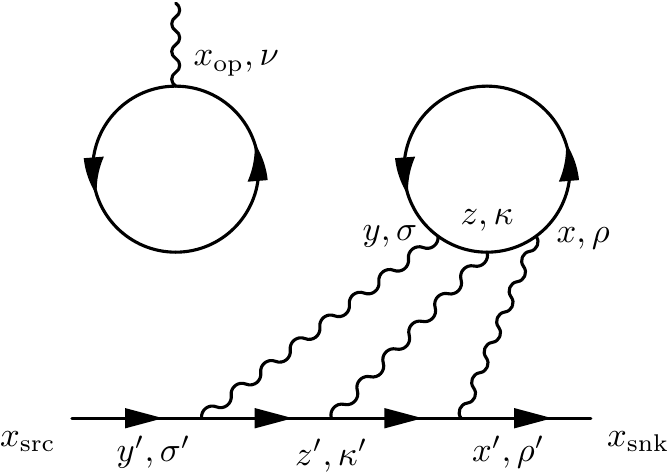}}
  \caption{\label{fig:dlbl-1}  Disconnected diagrams of order $m_s - m_l$. There are additional diagrams which can be obtained from permutation of the photon vertices on the muon line.}
\end{figure}

\begin{figure}[!htbp]
  \resizebox{0.4\columnwidth}{!}{\includegraphics{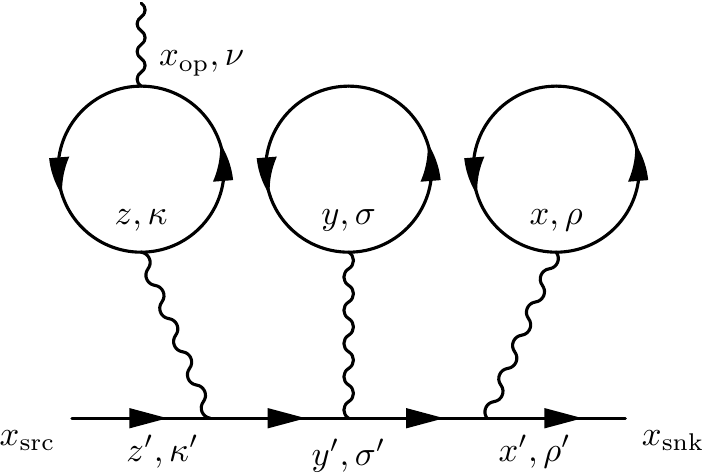}}
  \resizebox{0.4\columnwidth}{!}{\includegraphics{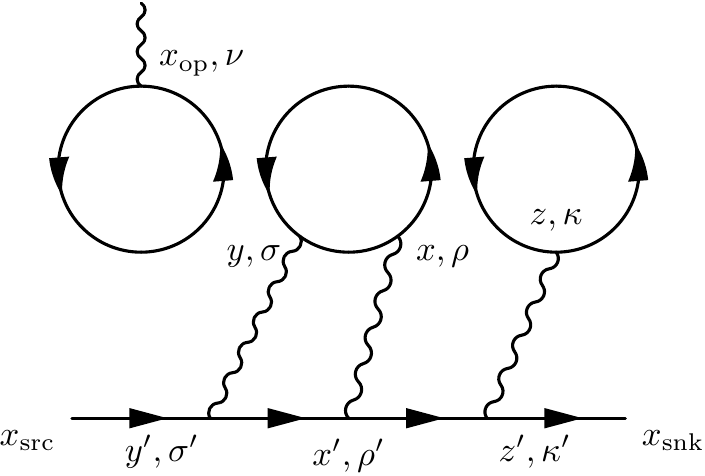}}
  \resizebox{0.5\columnwidth}{!}{\includegraphics{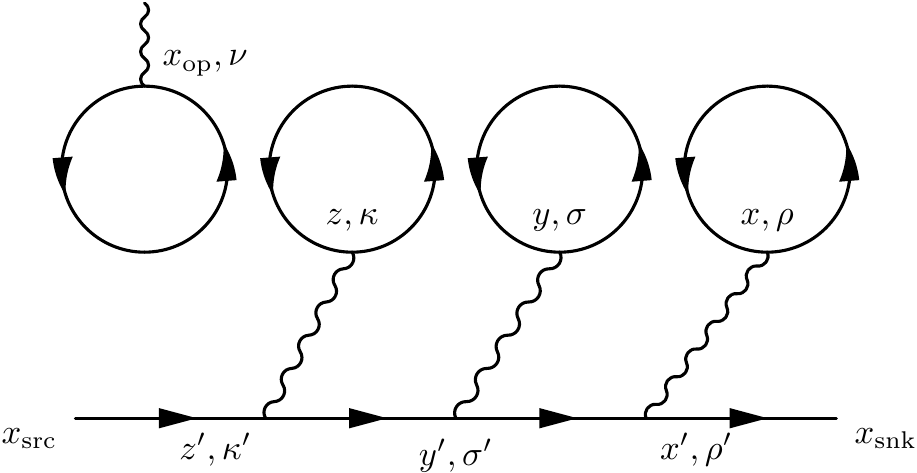}}
  \caption{\label{fig:dlbl-2}  Disconnected diagrams of order $(m_s - m_l)^2$ and higher. There are additional diagrams which can be obtained from permutation of the photon vertices on the muon line.}
\end{figure}

The complete set of HLbL diagrams include the connected diagrams in Fig.~\ref{fig:clbl} 
and the disconnected diagrams in Fig.~\ref{fig:dlbl-0}, 
\ref{fig:dlbl-1}, and \ref{fig:dlbl-2}. Only quark loops that
are directly connected to photons are drawn in the figures. This is because only
these quark propagators  need to be explicitly calculated on the lattice. 
%(Recall that each such propagator is the inverse of the Dirac operator computed in the background of the gauge field samples used to evaluate the path integral.)  
The effects of gluons
and other quark loops are included automatically through the evaluation of
these explicit quark propagators and the use of an unquenched gauge ensemble.  Although
there are many different types of disconnected diagrams, only one type, shown
in Fig.~\ref{fig:dlbl-0}, survives in the SU(3) limit.  The other diagrams, shown
in Figs.~\ref{fig:dlbl-1} and \ref{fig:dlbl-2}, vanish in SU(3) limit because they contain quark loops that couple only to one photon and the
sum of the charges of the $u$, $d$, $s$ quarks is zero. Also, because the strange
quark carries only $1 / 3$ of the electron charge, diagrams that are
suppressed by the difference between the strange and light quark masses are
suppressed by their charge factors too.

The first attempt using lattice QCD to compute the connected contribution to HLbL was made by Blum,
Chowdhury, Hayakawa, and Izubuchi {\cite{Blum:2014oka}}, which demonstrated the
possibility of performing such calculation. A series of improvements in methodology were
made in Ref.~{\cite{Blum:2015gfa}}, eliminating two sources of
systematic effects arising from the use of larger-than-physical electric charge and
non-zero momentum transfer. The methods presented in Ref.~{\cite{Blum:2015gfa}} also 
lead to a substantial reduction in the statistical noise making a direct lattice calculation with a
physical pion mass possible.  Here, we report the result of
the first connected HLbL lattice calculation with physical pion mass. In
addition to the connected HLbL calculation, we extended the methods of Ref.~{\cite{Blum:2015gfa}} and compute
the leading disconnected diagrams shown in Fig.~\ref{fig:dlbl-0}
using the same set of configurations.
This is the first disconnected HLbL calculation and the result suggests that
the leading order disconnected diagrams are quite important.

\section{Computational strategy}

We begin with the equation for the form factor $F_2^{\mathrm{cHLbL}}(0)$ derived in 
Ref.~\cite{Blum:2015gfa} for the moment method:
\begin{eqnarray}
 \frac{1}{2m} F_2^{\mathrm{cHLbL}} (q^2 = 0) \sigma^i_{s', s}
  &=& \sum_{r, \tilde{z}} \mathfrak{Z} \left( \frac{r}{2}, - \frac{r}{2},
  \tilde{z} \right)   \label{eq:f2-lbl-moment-short-z} \\
  && \hskip -1.4 in \times \sum_{\tilde{x}_{\text{op}}}  \frac{1}{2} \epsilon_{i, j,
  k} \left( \tilde{x}_{\text{op}} \right)_j \cdot i \bar{u}_{s'} (\vec{0})
  \mathcal{F}^C_k \left( \frac{r}{2}, - \frac{r}{2}, \tilde{z},
  \tilde{x}_{\text{op}} \right) u_s (\vec{0}), \nonumber
\end{eqnarray}
where the $\sigma^i_{s', s}$ are
the conventional Pauli matrices and the weight function $\mathfrak{Z}$ is defined by
\begin{eqnarray}
  \mathfrak{Z} (x, y, z) & = & \left\{ \begin{array}{ll}
  3 &
  \begin{array}{ll}
    \text{if}  & | x - y | < | x - z | \\
    \text{and} & | x - y | < | y - z |
  \end{array}\\
  3 / 2 &
  \begin{array}{ll}
    \text{if} & | x - y | = | x - z | < | y - z | \\
    \text{or} & | x - y | = | y - z | < | x - z |
  \end{array}\\
  1 & \text{if } | x - y | = | x - z | = | y - z |\\
  0 & \text{otherwise}
\end{array} \right. .  \label{eq:3-mult}
\end{eqnarray}
The integration variables in Eq.~\eqref{eq:f2-lbl-moment-short-z} are related to the 
coordinates in Fig.~\ref{fig:clbl} by the following equations: $r = x - y$, 
$\tilde{z} = z - (x +y) / 2$, $\tilde{x}_{\text{op}} = x_{\text{op}} - (x + y) / 2$. 
We compute the
summation over $r$ in Eq. \eqref{eq:f2-lbl-moment-short-z} by stochastically
sampling the points $x$ and $y$ while the sums over $\tilde{x}_{\text{op}}$ and
$\tilde{z}$ are performed exactly over the entire lattice.  The weight factor $\mathfrak{Z}$
exploits the symmetry of the integrand in Eq.~\eqref{eq:f2-lbl-moment-short-z} to insure 
that the exactly integrated point $z$ is at least as distant from both $x$ and $y$ as they
are from each other, resulting in a deterministic treatment of this difficult-to-sample, 
long-distance region.

The amplitude $\mathcal{F}^C_{\nu} \left( x, y, z, x_{\text{op}} \right)$ is obtained from the average of the function $\mathcal{F}_{\nu} \left( x, y, z, x_{\text{op}} \right)$ over the
three cyclic permutations of the positions $x$, $y$ and $z$ where
\begin{eqnarray}
  && \mathcal{F}_{\nu} \left( x, y, z, x_{\text{op}} \right) \\
  && \hskip 0.4 in = (- i e)^6 \,
  \mathcal{G}_{\rho, \sigma, \kappa} \left( x, y, z, x_{\text{snk}},
  x_{\text{src}} \right)
  \sum_{q = u, d, s} \left(\frac{e_q}{e}\right)^4 \nonumber \\
  && \hskip 0.5 in \times \Bigl\langle - \tmop{tr}
  \bigl[ \gamma_{\rho} S_q (x, z) \gamma_{\kappa} S_q (z, y) \gamma_{\sigma} 
\nonumber \\
&& \hskip 1.5 in 
 \cdot S_q \left( y, x_{\text{op}} \right) \gamma_{\nu} S_q \left( x_{\text{op}}, x
  \right) \bigr] \Bigr\rangle_{\text{QCD}} . \nonumber
  \label{eq:lbl-amp}
\end{eqnarray}
This equation expresses the connected HLbL amplitude as the average over QCD  gauge configurations of the trace of the product of four quark propagators ($S_q$) multiplied by a factor $ \mathcal{G}_{\rho,\sigma, \kappa}$, constructed from muon and photon propagators:
\begin{eqnarray}
  \mathcal{G}_{\rho, \sigma, \kappa} \left( x, y, z\right) 
   &&  = e^{m_{{\mu}} \left( t_{\text{snk}} - t_{\text{src}} \right)}  \label{eq:muon-line}\\
  && \hskip -0.75 in \times
  \sum_{x', y', z'} G_{\rho, \rho'} (x, x') G_{\sigma, \sigma'} (y, y')
  G_{\kappa, \kappa'} (z, z') \nonumber\\
  &&  \hskip -0.6 in \times\Biggl\{ \sum_{\vec{x}_{\text{snk}}, \vec{x}_{\text{src}}}
 \Bigl[S_{{\mu}} \left( x_{\text{snk}}, x' \right) \gamma_{\rho'} S_{{\mu}}
  (x', z') \gamma_{\kappa'} \nonumber \\
 && \hskip 0.7 in  \cdot S_{{\mu}} (z', y') \gamma_{\sigma'}
  S_{{\mu}} \left( y', x_{\text{src}} \right) \nonumber\\
   && \hskip 0.05 in +\quad  \text{five permutations of $x'$, $y'$ and $z'$} \Bigr]\Biggr\}. \nonumber
\end{eqnarray}
where $e_u / e = 2 / 3$, $e_d / e = e_s / e = - 1 / 3$.

We evaluate the muon propagators in Eq.~\eqref{eq:muon-line} using infinite-$L_s$, DWF on a lattice assigned 
the same lattice spacing and with the same lattice volume as the QCD gauge ensemble.  (Here $L_s$ is the extent of the fifth dimension for the DWF formalism.)  The 
photon propagators are evaluated in Feynman gauge and all modes with vanishing 
spatial momentum are omitted~\cite{Hayakawa:2008an}.  Because $\mathcal{G}$ involves a zero-mass
photon, finite volume effects are suppressed only by powers of the lattice
size. 

We also employ the moment method described above for the disconnected diagrams in Fig.~\ref{fig:dlbl-0} using
\begin{eqnarray}
  \frac{1}{2m}F_2^{\tmop{dHLbL}} (q^2 = 0) \sigma^i_{s', s}
  & = & \sum_{r, \tilde{x}} \sum_{\tilde{x}_{\text{op}}}  \frac{1}{2}
  \epsilon_{i, j, k} \left( \tilde{x}_{\text{op}} \right)_j  \\
  && \hskip -0.3 in \cdot i
  \bar{u}_{s'} (\vec{0}) \mathcal{F}^D_k \left( \tilde{x}, 0, r, r +
  \tilde{x}_{\text{op}} \right) u_s (\vec{0}). \nonumber
  \label{eq:f2-dlbl-moment}
\end{eqnarray}
The integration variables are related to the coordinates in Fig.~\ref{fig:dlbl-0} 
by the equations: $r = z - y$, $\tilde{x} = x - y$ and
$\tilde{x}_{\text{op}} = x_{\text{op}} - z$. As in the connected case, the
sum over $\tilde{x}$ and $\tilde{x}_{\tmop{op}}$ is performed over all lattice sites 
but the sum over $r$ is performed by stochastically sampling the points $z$ and $y$. The amplitude $\mathcal{F}^D_{\nu} \left( x, y, z, x_{\text{op}}\right)$ is given by:
\begin{eqnarray}
  \mathcal{F}^D_{\nu} \left( x, y, z, x_{\text{op}} \right) &=&
(- i e)^6
  e^{m_{{\mu}}  \left( t_{\text{snk}} - t_{\text{src}} \right)}
  \mathcal{G}_{\rho, \sigma, \kappa} \left( x, y, z \right)
  \label{eq:dlbl-amp}  \\
  && \hskip -0.8 in \times\left\langle \frac{1}{2} \Pi_{\nu, \kappa} \left(
  x_{\text{op}}, z \right)  \Bigl[ \Pi_{\rho, \sigma} (x, y) -
 \Pi_{\rho, \sigma}^{\text{avg}} (x - y) \Bigr] \right\rangle_{\text{QCD}}
  \nonumber
\end{eqnarray}
where
\begin{eqnarray}
  \Pi_{\rho, \sigma} (x, y) =  -\sum_q (e_q / e)^2 \tmop{Tr}
  \bigl[\gamma_{\rho} S_q (x, y) \gamma_{\sigma} S_q (y, x)\bigr].\mbox{~~}
\end{eqnarray}
The subtraction shown inside the square brackets in the second
line of Eq.~\eqref{eq:dlbl-amp} is performed only as a noise
reduction technique. It does not affect the central value provided the
subtraction term $\Pi_{\rho, \sigma}^{\text{avg}} (x - y)$ 
remains constant in the ensemble average.
This is a consequence of space-time reflection symmetry: $\left\langle \Pi_{\nu, \kappa} \left(
x_{\text{op}}, z \right) \right\rangle_{\text{QCD}} = \left\langle \Pi_{\nu, \kappa} \left( - x_{\text{op}}, - z \right) \right\rangle_{\text{QCD}}$. As a
result, the moment in Eq.~\eqref{eq:dlbl-amp} of a single factor of $ \left\langle \Pi_{\nu, \kappa} \left( x_{\text{op}}, z \right) \right\rangle$ vanishes:
\begin{eqnarray}
  \sum_{\tilde{x}_{\text{op}}} \frac{1}{2} \epsilon_{i, j, k} \left(
  \tilde{x}_{\text{op}} \right)_j  \left\langle \Pi_{k, \kappa} \left(
  \tilde{x}_{\text{op}}, 0 \right) \right\rangle_{\text{QCD}}  && \\
  && \hskip -1.6 in = \sum_{\tilde{x}_{\text{op}}} \frac{1}{2} \epsilon_{i, j, k} \left( -
  \tilde{x}_{\text{op}} \right)_j  \left\langle \Pi_{k, \kappa} \left( -
  \tilde{x}_{\text{op}}, 0 \right) \right\rangle_{\text{QCD}} = 0.
\nonumber
\end{eqnarray}
The ensemble average of the $\Pi$ function, $\langle \Pi_{\rho, \sigma} (x, y)
\rangle_{\text{QCD}}$, is a good choice for the subtraction term
$\Pi^{\text{avg}}_{\rho, \sigma} (x - y)$.  In our calculation, we set
$\Pi^{\text{avg}}_{\rho, \sigma} (x - y)$ to be the average of the
contractions of 32, uniformly-distributed, point-source propagators, all of
which are computed using a single configuration in our ensemble. That
configuration is not used elsewhere in the calculation.  

Note that a similar subtraction would be essential if the moment method 
were not being used, in order to avoid double counting a contribution that is 
conventionally included as two hadronic vacuum polarization corrections
to the $O(\alpha)$ QED contribution to $a_\mu$~\cite{Hayakawa:2015ntr}.

\section{Computational details}

The HLbL calculation is performed on the $48^3 \times 96$ physical-pion-mass
ensemble generated by the RBC and UKQCD collaborations~\cite{Blum:2014tka}.
The calculation is carried out on 65 configurations, separated by 20 molecular-dynamics time units.

In the connected-diagram calculation, for each configuration, we sample 112
short-distance point-pairs with $| r | \leqslant 5$ in lattice units, and 256
long-distance point-pairs with $| r | > 5$. The 112 short-distance distance
point-pairs cover all possible values of $r$ up to discrete symmetries on the 
lattice, which include reflections and $\pi / 2$ rotations. In fact, all $r$
with $| r | \leqslant 2$ are computed twice. For the long-distance
point-pairs, the probability of choosing one particular relative distance $r$ is 
$p(r) \propto \exp (- 0.01 | r |) / | r |^4$, an empirically suggested choice. 
The first point of all these
point-pairs is chosen independently, uniformly distributed over the
lattice. The second point is chosen according to the distribution
in $r$ described above.

In the disconnected-diagram calculation, for each configuration we randomly
choose 256 spheres, each of radius $6$, and 4 points are sampled
uniformly within each sphere. Duplication of the points is avoided in the sampling process.
Overall, 1024 points are sampled for each configuration.
Half of these points are also used to compute point-source, strange-quark propagators.
All the combinations of these points form $(1024 + 512)^2$ point-pairs
and all are used in the calculation.
This provides a very large number of point-pairs, sufficient to reduce the statistical error 
from the disconnected diagrams down to the level of the error from the connected diagrams.

The largest computational effort in this calculation is required to evaluate the
light quark propagators, making it important to use a method which 
gives results with sufficient accuracy at a minimum computational cost.  This is 
achieved here by using the AMA method~\cite{Blum:2012uh, Shintani:2014vja} 
detailed below.  In addition, we use the zM\"obius DWF variant~\cite{Mcglynn:2015uwh} 
to reduce 
$L_s$ from the value of 24, used in when generating the ensemble, to $L_s=10$, 
further accelerating the Dirac inversions. We use three accuracy levels in 
this AMA calculation: sloppy, median and exact.

Most of the light quark propagators are obtained using sloppy inversions.  
These are evaluated using the zM\"obius variant of DWF with $L_s = 10$, 
deflated using 2000 eigenvectors obtained from the Lanczos algorithm and 
200 single-precision, conjugate gradient (CG) iterations.  We extend 
some of these sloppy inversions with a defect-correction 
and deflation step, followed by 200 single-precision iterations to achieve
the median level of accuracy. Finally, the exact propagators are obtained from the
MADWF algorithm~\cite{Yin:2011np}, iterated until the results reach a precision 
of $10^{- 8}$ ({\it i.e.} the norm of the residual is $10^{-8}$ times 
smaller than the norm of the source).

For the strange quark propagators, no deflation is performed. The sloppy
inversions use the same zM\"obius Dirac operator with the strange quark 
mass and 300 single-precision iterations. The median inversions use an additional defect
correction step followed by 300 single-precision iterations. The exact inversions use
the unitary Mobius Dirac operator and then perform sufficient CG iterations that 
the propagators reach a $10^{- 8}$ precision.

In the connected-diagram calculation, only the light quark contribution is included.  
(The strange quark contribution will be substantially smaller because of both its 
heavier mass and an overall charge factor of 1/17.)  We compute
an additional 12 point-pairs with sloppy and median propagators and use the
correlated difference between the sloppy- and median-precision results to 
reduce the numerical error to that present in a median-precision calculation.
These 12 point pairs are sampled with probability $p (r) \propto \exp (- 0.07 | r |) /| r |^2$.
Finally, among these 12 point-pairs, 4 point-pairs are computed with exact 
propagators and the correlated difference between the median and
exact results is used to reduce the numerical error to the exact level. 
The corrections introduced at each step of this two-step AMA correction 
are 1\% and statistically consistent with zero.

In the disconnected-diagram calculation, both the light- and strange-quark 
contributions are included although the strange quark contributes less
than 5\%.   We use the same set of points for the
light- and strange-quark AMA corrections. From 512 points, 32 points are also
computed with median propagators.  Among these 32
points, 8 points are also computed with exact propagators.  The correlated
sloppy-median and median-exact differences are used to compute the correction. 
By making this correction we replace the unknown numerical error in the
sloppy-precision calculation with the known statistical error of the correction.

\section{Analysis and results}

Following the procedure described above we obtain the following results:
\begin{eqnarray}
  a_\mu^\text{cHLbL} &=& (11.60 \pm 0.96) \times 10^{- 10}  \label{eq:a-chlbl}\\
  a_\mu^\text{dHLbL} &=& (- 6.25 \pm 0.80) \times 10^{- 10}  \label{eq:a-dhlbl}\\
  a_\mu^\text{HLbL}   &=& (5.35 \pm 1.35) \times 10^{- 10},  \label{eq:a-hlbl}
\end{eqnarray}
where the errors shown are statistical only.  These results are obtained from a single
gauge ensemble with an inverse lattice spacing of $a^{-1}=1.730(4)$ and a spatial size of $L=5.476(12)$ fm.

\begin{figure}[!htbp]
  \resizebox{0.49\columnwidth}{!}{\includegraphics{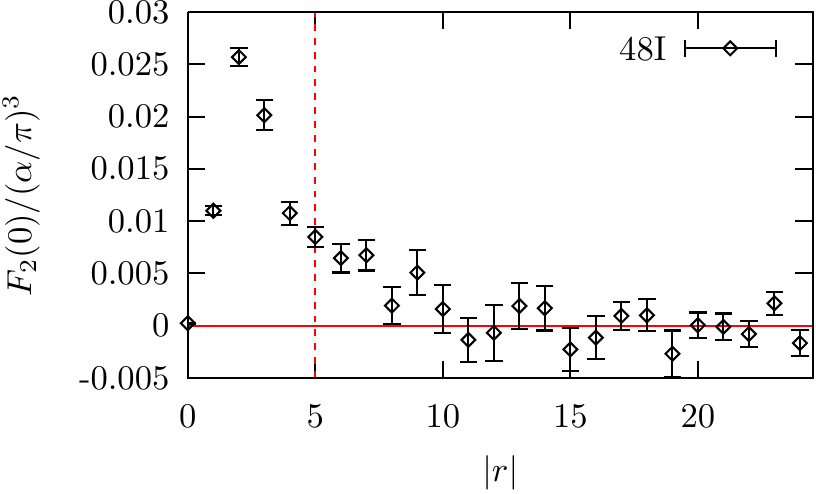}}
  \resizebox{0.49\columnwidth}{!}{\includegraphics{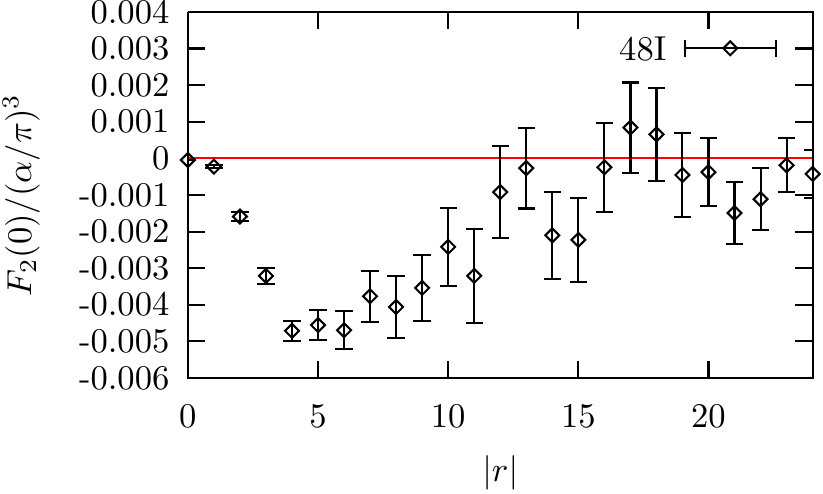}}
  \caption{\label{fig:histogram}Histograms of the contribution to $F_2$ from
  different separations $| r |$. The $i$th bin includes contribution from
  region where $i - 1 < | r | \leqslant i$. The sum of all bins gives
  the final result for $F_2$. Left: the cHLbL contribution where $r = x - y$. Right:
  the  dHLbL contribution where $r = z - y$.}
\end{figure}

Because the integration over $r$ is performed as the last step when evaluating 
Eqs.~\eqref{eq:f2-lbl-moment-short-z} and \eqref{eq:f2-dlbl-moment} by summing 
over the stochastically-sampled point-pairs, we can study the contribution to 
$F_2$ as a function of $r$ as shown in Fig.~\ref{fig:histogram}. 
From the left plot we can see that most of the 
connected-diagram contribution comes from a separation of $| r | \leqslant 10$ in
lattice units, while for the disconnected diagrams, the signal vanishes 
more slowly and its large-$r$ behavior is obscured by the noise.
The smaller, large-$r$ contribution seen for the connected diagrams comes partly 
from our use of the weight factor $\mathfrak{Z}$ in Eq.~\eqref{eq:f2-lbl-moment-short-z} to shift the contribution toward the short-distance region, a strategy not possible in the disconnected case.

The disconnected-diagram contribution is quite large and negative, which 
may be partly explained by the size of the $\pi^0$-pole contriubtions to the
disconnected parts of the amplitude~\cite{Bijnens:2016hgx,Jin:2016xxx}.

\section{Conclusion}

We have presented the first lattice calculation of the connected, hadronic
light-by-light contribution to the muon anomalous magnetic moment at physical
pion mass and the first lattice calculation of the leading disconnected
hadronic light-by-light contribution, also at physical pion mass. We find that
the disconnected diagrams contribute negatively and cancel approximately half 
of the connected contribution. While the combined result is much smaller
than that of most model calculations, we expect large finite-volume
and finite-lattice-spacing corrections, both of which were found to increase the 
result in our previous calculations using smaller lattice volumes~\cite{Blum:2015gfa}.  
Consequently, our lattice QCD result for the hadronic light-by-light scattering contribution to $a_\mu$ reported in Eq.~\eqref{eq:a-hlbl} should not be viewed as inconsistent with the result of model calculations.  However, this lattice result is completely independent from those model calculations, with unrelated systematic errors.  Therefore the calculation reported here makes it even more unlikely that the present discrepancy between the experimental result for $a_\mu$ and the prediction of the standard model might be completely explained by an error in the estimate of the hadronic light-by-light scattering contribution.

Since the largest finite-volume errors are expected
to arise from the QED part of the calculation, they may be reduced by 
performing only the QED part of the calculation in a larger or possibly infinite 
volume~\cite{Jin:2015bty, Green:2015mva, Asmussen:2016lse}.  This is an extension 
of the  calculation reported here which should be practical with current 
computational resources.  
The finite-lattice-spacing errors can be removed by performing the same
calculation on a $64^3 \times 128$ lattice with a smaller lattice 
spacing~\cite{Blum:2014tka}, which can then be combined with the present
calculation to extrapolate to vanishing lattice spacing, a calculation that is 
now underway.

\section{Ackownlegements}

We would like to thank our RBC and UKQCD collaborators for helpful discussions
and support. We would also like to thank RBRC for BG/Q computer time. The
BAGEL~\cite{Boyle:2009vp} library is used to compute all the propagators.
The CPS~\cite{Jung:2014ata} software package is also used in the calculation.
The computation is performed under the ALCC Program of the US DOE on the 
Blue Gene/Q (BG/Q) Mira computer at the Argonne Leadership Class Facility, a
DOE Office of Science Facility supported under Contract De-AC02-06CH11357.
T.B. is supported by US DOE grant DE-FG02-92ER40716.
N.H.C. and L.C.J. are supported in part by US DOE grant \#DE-SC0011941.
M.H. is supported in part by Japan Grants-in-Aid for Scientific Research, No.16K05317.
T.I., C.J., and C.L. are supported in part by US DOE Contract \#AC-02-98CH10886(BNL).
T.I. is supported in part by the Japanese Ministry of Education Grant-in-Aid, No. 26400261.
CL acknowledges support through a DOE Office of Science Early Career Award.

\bibliography{ref}

\begin{thebibliography}{40}%
\makeatletter
\providecommand \@ifxundefined [1]{%
 \@ifx{#1\undefined}
}%
\providecommand \@ifnum [1]{%
 \ifnum #1\expandafter \@firstoftwo
 \else \expandafter \@secondoftwo
 \fi
}%
\providecommand \@ifx [1]{%
 \ifx #1\expandafter \@firstoftwo
 \else \expandafter \@secondoftwo
 \fi
}%
\providecommand \natexlab [1]{#1}%
\providecommand \enquote  [1]{``#1''}%
\providecommand \bibnamefont  [1]{#1}%
\providecommand \bibfnamefont [1]{#1}%
\providecommand \citenamefont [1]{#1}%
\providecommand \href@noop [0]{\@secondoftwo}%
\providecommand \href [0]{\begingroup \@sanitize@url \@href}%
\providecommand \@href[1]{\@@startlink{#1}\@@href}%
\providecommand \@@href[1]{\endgroup#1\@@endlink}%
\providecommand \@sanitize@url [0]{\catcode `\\12\catcode `\$12\catcode
  `\&12\catcode `\#12\catcode `\^12\catcode `\_12\catcode `\%12\relax}%
\providecommand \@@startlink[1]{}%
\providecommand \@@endlink[0]{}%
\providecommand \url  [0]{\begingroup\@sanitize@url \@url }%
\providecommand \@url [1]{\endgroup\@href {#1}{\urlprefix }}%
\providecommand \urlprefix  [0]{URL }%
\providecommand \Eprint [0]{\href }%
\providecommand \doibase [0]{http://dx.doi.org/}%
\providecommand \selectlanguage [0]{\@gobble}%
\providecommand \bibinfo  [0]{\@secondoftwo}%
\providecommand \bibfield  [0]{\@secondoftwo}%
\providecommand \translation [1]{[#1]}%
\providecommand \BibitemOpen [0]{}%
\providecommand \bibitemStop [0]{}%
\providecommand \bibitemNoStop [0]{.\EOS\space}%
\providecommand \EOS [0]{\spacefactor3000\relax}%
\providecommand \BibitemShut  [1]{\csname bibitem#1\endcsname}%
\let\auto@bib@innerbib\@empty
%</preamble>
\bibitem [{\citenamefont {Jegerlehner}\ and\ \citenamefont
  {Nyffeler}(2009)}]{Jegerlehner:2009ry}%
  \BibitemOpen
  \bibfield  {author} {\bibinfo {author} {\bibfnamefont {F.}~\bibnamefont
  {Jegerlehner}}\ and\ \bibinfo {author} {\bibfnamefont {A.}~\bibnamefont
  {Nyffeler}},\ }\href {\doibase 10.1016/j.physrep.2009.04.003} {\bibfield
  {journal} {\bibinfo  {journal} {Phys. Rept.}\ }\textbf {\bibinfo {volume}
  {477}},\ \bibinfo {pages} {1} (\bibinfo {year} {2009})},\ \Eprint
  {http://arxiv.org/abs/0902.3360} {arXiv:0902.3360 [hep-ph]} \BibitemShut
  {NoStop}%
%%CITATION = ARXIV:0902.3360;%%
\bibitem [{\citenamefont {Schwinger}(1948)}]{Schwinger:1948iu}%
  \BibitemOpen
  \bibfield  {author} {\bibinfo {author} {\bibfnamefont {J.~S.}\ \bibnamefont
  {Schwinger}},\ }\href {\doibase 10.1103/PhysRev.73.416} {\bibfield  {journal}
  {\bibinfo  {journal} {Phys. Rev.}\ }\textbf {\bibinfo {volume} {73}},\
  \bibinfo {pages} {416} (\bibinfo {year} {1948})}\BibitemShut {NoStop}%
%%CITATION = PHRVA,73,416;%%
\bibitem [{\citenamefont {Bennett}\ \emph {et~al.}(2006)\citenamefont {Bennett}
  \emph {et~al.}}]{Bennett:2006fi}%
  \BibitemOpen
  \bibfield  {author} {\bibinfo {author} {\bibfnamefont {G.}~\bibnamefont
  {Bennett}} \emph {et~al.} (\bibinfo {collaboration} {Muon G-2
  Collaboration}),\ }\href {\doibase 10.1103/PhysRevD.73.072003} {\bibfield
  {journal} {\bibinfo  {journal} {Phys.Rev.}\ }\textbf {\bibinfo {volume}
  {D73}},\ \bibinfo {pages} {072003} (\bibinfo {year} {2006})},\ \Eprint
  {http://arxiv.org/abs/hep-ex/0602035} {arXiv:hep-ex/0602035 [hep-ex]}
  \BibitemShut {NoStop}%
%%CITATION = HEP-EX/0602035;%%
\bibitem [{\citenamefont {Carey}\ \emph {et~al.}(2009)\citenamefont {Carey}
  \emph {et~al.}}]{Carey:2009zzb}%
  \BibitemOpen
  \bibfield  {author} {\bibinfo {author} {\bibfnamefont {R.~M.}\ \bibnamefont
  {Carey}} \emph {et~al.},\ }\href@noop {} {\  (\bibinfo {year}
  {2009})}\BibitemShut {NoStop}%
%%CITATION = FERMILAB-PROPOSAL-0989;%%
\bibitem [{\citenamefont {Mibe}(2010)}]{Mibe:2010zz}%
  \BibitemOpen
  \bibfield  {author} {\bibinfo {author} {\bibfnamefont {T.}~\bibnamefont
  {Mibe}} (\bibinfo {collaboration} {J-PARC g-2}),\ }\bibfield  {booktitle}
  {\emph {\bibinfo {booktitle} {{Proceedings, 6th International Workshop on
  e+e- Collisions from Phi to Psi (PHIPSI09): Beijing, China, October 13-16,
  2009}}},\ }\href {\doibase 10.1088/1674-1137/34/6/022} {\bibfield  {journal}
  {\bibinfo  {journal} {Chin. Phys.}\ }\textbf {\bibinfo {volume} {C34}},\
  \bibinfo {pages} {745} (\bibinfo {year} {2010})}\BibitemShut {NoStop}%
%%CITATION = CHPHD,C34,745;%%
\bibitem [{\citenamefont {Aoyama}\ \emph {et~al.}(2012)\citenamefont {Aoyama},
  \citenamefont {Hayakawa}, \citenamefont {Kinoshita},\ and\ \citenamefont
  {Nio}}]{Aoyama:2012wk}%
  \BibitemOpen
  \bibfield  {author} {\bibinfo {author} {\bibfnamefont {T.}~\bibnamefont
  {Aoyama}}, \bibinfo {author} {\bibfnamefont {M.}~\bibnamefont {Hayakawa}},
  \bibinfo {author} {\bibfnamefont {T.}~\bibnamefont {Kinoshita}}, \ and\
  \bibinfo {author} {\bibfnamefont {M.}~\bibnamefont {Nio}},\ }\href {\doibase
  10.1103/PhysRevLett.109.111808} {\bibfield  {journal} {\bibinfo  {journal}
  {Phys.Rev.Lett.}\ }\textbf {\bibinfo {volume} {109}},\ \bibinfo {pages}
  {111808} (\bibinfo {year} {2012})},\ \Eprint {http://arxiv.org/abs/1205.5370}
  {arXiv:1205.5370 [hep-ph]} \BibitemShut {NoStop}%
%%CITATION = ARXIV:1205.5370;%%
\bibitem [{\citenamefont {Gnendiger}\ \emph {et~al.}(2013)\citenamefont
  {Gnendiger}, \citenamefont {Stockinger},\ and\ \citenamefont
  {Stockinger-Kim}}]{Gnendiger:2013pva}%
  \BibitemOpen
  \bibfield  {author} {\bibinfo {author} {\bibfnamefont {C.}~\bibnamefont
  {Gnendiger}}, \bibinfo {author} {\bibfnamefont {D.}~\bibnamefont
  {Stockinger}}, \ and\ \bibinfo {author} {\bibfnamefont {H.}~\bibnamefont
  {Stockinger-Kim}},\ }\href {\doibase 10.1103/PhysRevD.88.053005} {\bibfield
  {journal} {\bibinfo  {journal} {Phys.Rev.}\ }\textbf {\bibinfo {volume}
  {D88}},\ \bibinfo {pages} {053005} (\bibinfo {year} {2013})},\ \Eprint
  {http://arxiv.org/abs/1306.5546} {arXiv:1306.5546 [hep-ph]} \BibitemShut
  {NoStop}%
%%CITATION = ARXIV:1306.5546;%%
\bibitem [{\citenamefont {Davier}\ \emph {et~al.}(2011)\citenamefont {Davier},
  \citenamefont {Hoecker}, \citenamefont {Malaescu},\ and\ \citenamefont
  {Zhang}}]{Davier:2010nc}%
  \BibitemOpen
  \bibfield  {author} {\bibinfo {author} {\bibfnamefont {M.}~\bibnamefont
  {Davier}}, \bibinfo {author} {\bibfnamefont {A.}~\bibnamefont {Hoecker}},
  \bibinfo {author} {\bibfnamefont {B.}~\bibnamefont {Malaescu}}, \ and\
  \bibinfo {author} {\bibfnamefont {Z.}~\bibnamefont {Zhang}},\ }\href
  {\doibase 10.1140/epjc/s10052-012-1874-8, 10.1140/epjc/s10052-010-1515-z}
  {\bibfield  {journal} {\bibinfo  {journal} {Eur.Phys.J.}\ }\textbf {\bibinfo
  {volume} {C71}},\ \bibinfo {pages} {1515} (\bibinfo {year} {2011})},\ \Eprint
  {http://arxiv.org/abs/1010.4180} {arXiv:1010.4180 [hep-ph]} \BibitemShut
  {NoStop}%
%%CITATION = ARXIV:1010.4180;%%
\bibitem [{\citenamefont {Hagiwara}\ \emph {et~al.}(2011)\citenamefont
  {Hagiwara}, \citenamefont {Liao}, \citenamefont {Martin}, \citenamefont
  {Nomura},\ and\ \citenamefont {Teubner}}]{Hagiwara:2011af}%
  \BibitemOpen
  \bibfield  {author} {\bibinfo {author} {\bibfnamefont {K.}~\bibnamefont
  {Hagiwara}}, \bibinfo {author} {\bibfnamefont {R.}~\bibnamefont {Liao}},
  \bibinfo {author} {\bibfnamefont {A.~D.}\ \bibnamefont {Martin}}, \bibinfo
  {author} {\bibfnamefont {D.}~\bibnamefont {Nomura}}, \ and\ \bibinfo {author}
  {\bibfnamefont {T.}~\bibnamefont {Teubner}},\ }\href {\doibase
  10.1088/0954-3899/38/8/085003} {\bibfield  {journal} {\bibinfo  {journal}
  {J.Phys.}\ }\textbf {\bibinfo {volume} {G38}},\ \bibinfo {pages} {085003}
  (\bibinfo {year} {2011})},\ \Eprint {http://arxiv.org/abs/1105.3149}
  {arXiv:1105.3149 [hep-ph]} \BibitemShut {NoStop}%
%%CITATION = ARXIV:1105.3149;%%
\bibitem [{\citenamefont {Grange}\ \emph {et~al.}(2015)\citenamefont {Grange}
  \emph {et~al.}}]{Grange:2015fou}%
  \BibitemOpen
  \bibfield  {author} {\bibinfo {author} {\bibfnamefont {J.}~\bibnamefont
  {Grange}} \emph {et~al.} (\bibinfo {collaboration} {Muon g-2}),\ }\href@noop
  {} {\  (\bibinfo {year} {2015})},\ \Eprint {http://arxiv.org/abs/1501.06858}
  {arXiv:1501.06858 [physics.ins-det]} \BibitemShut {NoStop}%
%%CITATION = ARXIV:1501.06858;%%
\bibitem [{\citenamefont {Bernecker}\ and\ \citenamefont
  {Meyer}(2011)}]{Bernecker:2011gh}%
  \BibitemOpen
  \bibfield  {author} {\bibinfo {author} {\bibfnamefont {D.}~\bibnamefont
  {Bernecker}}\ and\ \bibinfo {author} {\bibfnamefont {H.~B.}\ \bibnamefont
  {Meyer}},\ }\href {\doibase 10.1140/epja/i2011-11148-6} {\bibfield  {journal}
  {\bibinfo  {journal} {Eur. Phys. J.}\ }\textbf {\bibinfo {volume} {A47}},\
  \bibinfo {pages} {148} (\bibinfo {year} {2011})},\ \Eprint
  {http://arxiv.org/abs/1107.4388} {arXiv:1107.4388 [hep-lat]} \BibitemShut
  {NoStop}%
%%CITATION = ARXIV:1107.4388;%%
\bibitem [{\citenamefont {Feng}\ \emph {et~al.}(2013)\citenamefont {Feng},
  \citenamefont {Hashimoto}, \citenamefont {Hotzel}, \citenamefont {Jansen},
  \citenamefont {Petschlies},\ and\ \citenamefont {Renner}}]{Feng:2013xsa}%
  \BibitemOpen
  \bibfield  {author} {\bibinfo {author} {\bibfnamefont {X.}~\bibnamefont
  {Feng}}, \bibinfo {author} {\bibfnamefont {S.}~\bibnamefont {Hashimoto}},
  \bibinfo {author} {\bibfnamefont {G.}~\bibnamefont {Hotzel}}, \bibinfo
  {author} {\bibfnamefont {K.}~\bibnamefont {Jansen}}, \bibinfo {author}
  {\bibfnamefont {M.}~\bibnamefont {Petschlies}}, \ and\ \bibinfo {author}
  {\bibfnamefont {D.~B.}\ \bibnamefont {Renner}},\ }\href {\doibase
  10.1103/PhysRevD.88.034505} {\bibfield  {journal} {\bibinfo  {journal} {Phys.
  Rev.}\ }\textbf {\bibinfo {volume} {D88}},\ \bibinfo {pages} {034505}
  (\bibinfo {year} {2013})},\ \Eprint {http://arxiv.org/abs/1305.5878}
  {arXiv:1305.5878 [hep-lat]} \BibitemShut {NoStop}%
%%CITATION = ARXIV:1305.5878;%%
\bibitem [{\citenamefont {Blum}\ \emph
  {et~al.}(2016{\natexlab{a}})\citenamefont {Blum}, \citenamefont {Boyle},
  \citenamefont {Izubuchi}, \citenamefont {Jin}, \citenamefont {Juettner},
  \citenamefont {Lehner}, \citenamefont {Maltman}, \citenamefont {Marinkovic},
  \citenamefont {Portelli},\ and\ \citenamefont {Spraggs}}]{Blum:2015you}%
  \BibitemOpen
  \bibfield  {author} {\bibinfo {author} {\bibfnamefont {T.}~\bibnamefont
  {Blum}}, \bibinfo {author} {\bibfnamefont {P.~A.}\ \bibnamefont {Boyle}},
  \bibinfo {author} {\bibfnamefont {T.}~\bibnamefont {Izubuchi}}, \bibinfo
  {author} {\bibfnamefont {L.}~\bibnamefont {Jin}}, \bibinfo {author}
  {\bibfnamefont {A.}~\bibnamefont {Juettner}}, \bibinfo {author}
  {\bibfnamefont {C.}~\bibnamefont {Lehner}}, \bibinfo {author} {\bibfnamefont
  {K.}~\bibnamefont {Maltman}}, \bibinfo {author} {\bibfnamefont
  {M.}~\bibnamefont {Marinkovic}}, \bibinfo {author} {\bibfnamefont
  {A.}~\bibnamefont {Portelli}}, \ and\ \bibinfo {author} {\bibfnamefont
  {M.}~\bibnamefont {Spraggs}},\ }\href {\doibase
  10.1103/PhysRevLett.116.232002} {\bibfield  {journal} {\bibinfo  {journal}
  {Phys. Rev. Lett.}\ }\textbf {\bibinfo {volume} {116}},\ \bibinfo {pages}
  {232002} (\bibinfo {year} {2016}{\natexlab{a}})},\ \Eprint
  {http://arxiv.org/abs/1512.09054} {arXiv:1512.09054 [hep-lat]} \BibitemShut
  {NoStop}%
%%CITATION = ARXIV:1512.09054;%%
\bibitem [{\citenamefont {Blum}\ \emph
  {et~al.}(2016{\natexlab{b}})\citenamefont {Blum} \emph
  {et~al.}}]{Blum:2016xpd}%
  \BibitemOpen
  \bibfield  {author} {\bibinfo {author} {\bibfnamefont {T.}~\bibnamefont
  {Blum}} \emph {et~al.} (\bibinfo {collaboration} {RBC/UKQCD}),\ }\href
  {\doibase 10.1007/JHEP04(2016)063} {\bibfield  {journal} {\bibinfo  {journal}
  {JHEP}\ }\textbf {\bibinfo {volume} {04}},\ \bibinfo {pages} {063} (\bibinfo
  {year} {2016}{\natexlab{b}})},\ \Eprint {http://arxiv.org/abs/1602.01767}
  {arXiv:1602.01767 [hep-lat]} \BibitemShut {NoStop}%
%%CITATION = ARXIV:1602.01767;%%
\bibitem [{\citenamefont {Chakraborty}\ \emph {et~al.}(2016)\citenamefont
  {Chakraborty}, \citenamefont {Davies}, \citenamefont {de~Oliviera},
  \citenamefont {Koponen},\ and\ \citenamefont {Lepage}}]{Chakraborty:2016mwy}%
  \BibitemOpen
  \bibfield  {author} {\bibinfo {author} {\bibfnamefont {B.}~\bibnamefont
  {Chakraborty}}, \bibinfo {author} {\bibfnamefont {C.~T.~H.}\ \bibnamefont
  {Davies}}, \bibinfo {author} {\bibfnamefont {P.~G.}\ \bibnamefont
  {de~Oliviera}}, \bibinfo {author} {\bibfnamefont {J.}~\bibnamefont
  {Koponen}}, \ and\ \bibinfo {author} {\bibfnamefont {G.~P.}\ \bibnamefont
  {Lepage}},\ }\href@noop {} {\  (\bibinfo {year} {2016})},\ \Eprint
  {http://arxiv.org/abs/1601.03071} {arXiv:1601.03071 [hep-lat]} \BibitemShut
  {NoStop}%
%%CITATION = ARXIV:1601.03071;%%
\bibitem [{\citenamefont {Prades}\ \emph {et~al.}(2009)\citenamefont {Prades},
  \citenamefont {de~Rafael},\ and\ \citenamefont {Vainshtein}}]{Prades:2009tw}%
  \BibitemOpen
  \bibfield  {author} {\bibinfo {author} {\bibfnamefont {J.}~\bibnamefont
  {Prades}}, \bibinfo {author} {\bibfnamefont {E.}~\bibnamefont {de~Rafael}}, \
  and\ \bibinfo {author} {\bibfnamefont {A.}~\bibnamefont {Vainshtein}},\
  }\href@noop {} {\  (\bibinfo {year} {2009})},\ \Eprint
  {http://arxiv.org/abs/0901.0306} {arXiv:0901.0306 [hep-ph]} \BibitemShut
  {NoStop}%
%%CITATION = ARXIV:0901.0306;%%
\bibitem [{Ben(2014)}]{Benayoun:2014tra}%
  \BibitemOpen
  \href {http://inspirehep.net/record/1306493/files/arXiv:1407.4021.pdf} {\emph
  {\bibinfo {title} {{Hadronic contributions to the muon anomalous magnetic
  moment Workshop. $(g-2)_{\mu}$: Quo vadis? Workshop. Mini proceedings}}}}\
  (\bibinfo {year} {2014})\ \Eprint {http://arxiv.org/abs/1407.4021}
  {arXiv:1407.4021 [hep-ph]} \BibitemShut {NoStop}%
%%CITATION = ARXIV:1407.4021;%%
\bibitem [{\citenamefont {Green}\ \emph {et~al.}(2015)\citenamefont {Green},
  \citenamefont {Gryniuk}, \citenamefont {von Hippel}, \citenamefont {Meyer},\
  and\ \citenamefont {Pascalutsa}}]{Green:2015sra}%
  \BibitemOpen
  \bibfield  {author} {\bibinfo {author} {\bibfnamefont {J.}~\bibnamefont
  {Green}}, \bibinfo {author} {\bibfnamefont {O.}~\bibnamefont {Gryniuk}},
  \bibinfo {author} {\bibfnamefont {G.}~\bibnamefont {von Hippel}}, \bibinfo
  {author} {\bibfnamefont {H.~B.}\ \bibnamefont {Meyer}}, \ and\ \bibinfo
  {author} {\bibfnamefont {V.}~\bibnamefont {Pascalutsa}},\ }\href {\doibase
  10.1103/PhysRevLett.115.222003} {\bibfield  {journal} {\bibinfo  {journal}
  {Phys. Rev. Lett.}\ }\textbf {\bibinfo {volume} {115}},\ \bibinfo {pages}
  {222003} (\bibinfo {year} {2015})},\ \Eprint
  {http://arxiv.org/abs/1507.01577} {arXiv:1507.01577 [hep-lat]} \BibitemShut
  {NoStop}%
%%CITATION = ARXIV:1507.01577;%%
\bibitem [{\citenamefont {Colangelo}\ \emph
  {et~al.}(2014{\natexlab{a}})\citenamefont {Colangelo}, \citenamefont
  {Hoferichter}, \citenamefont {Procura},\ and\ \citenamefont
  {Stoffer}}]{Colangelo:2014dfa}%
  \BibitemOpen
  \bibfield  {author} {\bibinfo {author} {\bibfnamefont {G.}~\bibnamefont
  {Colangelo}}, \bibinfo {author} {\bibfnamefont {M.}~\bibnamefont
  {Hoferichter}}, \bibinfo {author} {\bibfnamefont {M.}~\bibnamefont
  {Procura}}, \ and\ \bibinfo {author} {\bibfnamefont {P.}~\bibnamefont
  {Stoffer}},\ }\href {\doibase 10.1007/JHEP09(2014)091} {\bibfield  {journal}
  {\bibinfo  {journal} {JHEP}\ }\textbf {\bibinfo {volume} {09}},\ \bibinfo
  {pages} {091} (\bibinfo {year} {2014}{\natexlab{a}})},\ \Eprint
  {http://arxiv.org/abs/1402.7081} {arXiv:1402.7081 [hep-ph]} \BibitemShut
  {NoStop}%
%%CITATION = ARXIV:1402.7081;%%
\bibitem [{\citenamefont {Colangelo}\ \emph
  {et~al.}(2014{\natexlab{b}})\citenamefont {Colangelo}, \citenamefont
  {Hoferichter}, \citenamefont {Kubis}, \citenamefont {Procura},\ and\
  \citenamefont {Stoffer}}]{Colangelo:2014pva}%
  \BibitemOpen
  \bibfield  {author} {\bibinfo {author} {\bibfnamefont {G.}~\bibnamefont
  {Colangelo}}, \bibinfo {author} {\bibfnamefont {M.}~\bibnamefont
  {Hoferichter}}, \bibinfo {author} {\bibfnamefont {B.}~\bibnamefont {Kubis}},
  \bibinfo {author} {\bibfnamefont {M.}~\bibnamefont {Procura}}, \ and\
  \bibinfo {author} {\bibfnamefont {P.}~\bibnamefont {Stoffer}},\ }\href
  {\doibase 10.1016/j.physletb.2014.09.021} {\bibfield  {journal} {\bibinfo
  {journal} {Phys. Lett.}\ }\textbf {\bibinfo {volume} {B738}},\ \bibinfo
  {pages} {6} (\bibinfo {year} {2014}{\natexlab{b}})},\ \Eprint
  {http://arxiv.org/abs/1408.2517} {arXiv:1408.2517 [hep-ph]} \BibitemShut
  {NoStop}%
%%CITATION = ARXIV:1408.2517;%%
\bibitem [{\citenamefont {Pauk}\ and\ \citenamefont
  {Vanderhaeghen}(2014{\natexlab{a}})}]{Pauk:2014jza}%
  \BibitemOpen
  \bibfield  {author} {\bibinfo {author} {\bibfnamefont {V.}~\bibnamefont
  {Pauk}}\ and\ \bibinfo {author} {\bibfnamefont {M.}~\bibnamefont
  {Vanderhaeghen}},\ }\href@noop {} {\  (\bibinfo {year}
  {2014}{\natexlab{a}})},\ \Eprint {http://arxiv.org/abs/1403.7503}
  {arXiv:1403.7503 [hep-ph]} \BibitemShut {NoStop}%
%%CITATION = ARXIV:1403.7503;%%
\bibitem [{\citenamefont {Pauk}\ and\ \citenamefont
  {Vanderhaeghen}(2014{\natexlab{b}})}]{Pauk:2014rfa}%
  \BibitemOpen
  \bibfield  {author} {\bibinfo {author} {\bibfnamefont {V.}~\bibnamefont
  {Pauk}}\ and\ \bibinfo {author} {\bibfnamefont {M.}~\bibnamefont
  {Vanderhaeghen}},\ }\href {\doibase 10.1103/PhysRevD.90.113012} {\bibfield
  {journal} {\bibinfo  {journal} {Phys. Rev.}\ }\textbf {\bibinfo {volume}
  {D90}},\ \bibinfo {pages} {113012} (\bibinfo {year} {2014}{\natexlab{b}})},\
  \Eprint {http://arxiv.org/abs/1409.0819} {arXiv:1409.0819 [hep-ph]}
  \BibitemShut {NoStop}%
%%CITATION = ARXIV:1409.0819;%%
\bibitem [{\citenamefont {Colangelo}\ \emph {et~al.}(2015)\citenamefont
  {Colangelo}, \citenamefont {Hoferichter}, \citenamefont {Procura},\ and\
  \citenamefont {Stoffer}}]{Colangelo:2015ama}%
  \BibitemOpen
  \bibfield  {author} {\bibinfo {author} {\bibfnamefont {G.}~\bibnamefont
  {Colangelo}}, \bibinfo {author} {\bibfnamefont {M.}~\bibnamefont
  {Hoferichter}}, \bibinfo {author} {\bibfnamefont {M.}~\bibnamefont
  {Procura}}, \ and\ \bibinfo {author} {\bibfnamefont {P.}~\bibnamefont
  {Stoffer}},\ }\href {\doibase 10.1007/JHEP09(2015)074} {\bibfield  {journal}
  {\bibinfo  {journal} {JHEP}\ }\textbf {\bibinfo {volume} {09}},\ \bibinfo
  {pages} {074} (\bibinfo {year} {2015})},\ \Eprint
  {http://arxiv.org/abs/1506.01386} {arXiv:1506.01386 [hep-ph]} \BibitemShut
  {NoStop}%
%%CITATION = ARXIV:1506.01386;%%
\bibitem [{\citenamefont {Nyffeler}(2016)}]{Nyffeler:2016gnb}%
  \BibitemOpen
  \bibfield  {author} {\bibinfo {author} {\bibfnamefont {A.}~\bibnamefont
  {Nyffeler}},\ }\href@noop {} {\  (\bibinfo {year} {2016})},\ \Eprint
  {http://arxiv.org/abs/1602.03398} {arXiv:1602.03398 [hep-ph]} \BibitemShut
  {NoStop}%
%%CITATION = ARXIV:1602.03398;%%
\bibitem [{\citenamefont {Blum}\ \emph {et~al.}(2015)\citenamefont {Blum},
  \citenamefont {Chowdhury}, \citenamefont {Hayakawa},\ and\ \citenamefont
  {Izubuchi}}]{Blum:2014oka}%
  \BibitemOpen
  \bibfield  {author} {\bibinfo {author} {\bibfnamefont {T.}~\bibnamefont
  {Blum}}, \bibinfo {author} {\bibfnamefont {S.}~\bibnamefont {Chowdhury}},
  \bibinfo {author} {\bibfnamefont {M.}~\bibnamefont {Hayakawa}}, \ and\
  \bibinfo {author} {\bibfnamefont {T.}~\bibnamefont {Izubuchi}},\ }\href
  {\doibase 10.1103/PhysRevLett.114.012001} {\bibfield  {journal} {\bibinfo
  {journal} {Phys.Rev.Lett.}\ }\textbf {\bibinfo {volume} {114}},\ \bibinfo
  {pages} {012001} (\bibinfo {year} {2015})},\ \Eprint
  {http://arxiv.org/abs/1407.2923} {arXiv:1407.2923 [hep-lat]} \BibitemShut
  {NoStop}%
%%CITATION = ARXIV:1407.2923;%%
\bibitem [{\citenamefont {Blum}\ \emph
  {et~al.}(2016{\natexlab{c}})\citenamefont {Blum}, \citenamefont {Christ},
  \citenamefont {Hayakawa}, \citenamefont {Izubuchi}, \citenamefont {Jin},\
  and\ \citenamefont {Lehner}}]{Blum:2015gfa}%
  \BibitemOpen
  \bibfield  {author} {\bibinfo {author} {\bibfnamefont {T.}~\bibnamefont
  {Blum}}, \bibinfo {author} {\bibfnamefont {N.}~\bibnamefont {Christ}},
  \bibinfo {author} {\bibfnamefont {M.}~\bibnamefont {Hayakawa}}, \bibinfo
  {author} {\bibfnamefont {T.}~\bibnamefont {Izubuchi}}, \bibinfo {author}
  {\bibfnamefont {L.}~\bibnamefont {Jin}}, \ and\ \bibinfo {author}
  {\bibfnamefont {C.}~\bibnamefont {Lehner}},\ }\href {\doibase
  10.1103/PhysRevD.93.014503} {\bibfield  {journal} {\bibinfo  {journal} {Phys.
  Rev.}\ }\textbf {\bibinfo {volume} {D93}},\ \bibinfo {pages} {014503}
  (\bibinfo {year} {2016}{\natexlab{c}})},\ \Eprint
  {http://arxiv.org/abs/1510.07100} {arXiv:1510.07100 [hep-lat]} \BibitemShut
  {NoStop}%
%%CITATION = ARXIV:1510.07100;%%
\bibitem [{\citenamefont {Hayakawa}\ and\ \citenamefont
  {Uno}(2008)}]{Hayakawa:2008an}%
  \BibitemOpen
  \bibfield  {author} {\bibinfo {author} {\bibfnamefont {M.}~\bibnamefont
  {Hayakawa}}\ and\ \bibinfo {author} {\bibfnamefont {S.}~\bibnamefont {Uno}},\
  }\href {\doibase 10.1143/PTP.120.413} {\bibfield  {journal} {\bibinfo
  {journal} {Prog. Theor. Phys.}\ }\textbf {\bibinfo {volume} {120}},\ \bibinfo
  {pages} {413} (\bibinfo {year} {2008})},\ \Eprint
  {http://arxiv.org/abs/0804.2044} {arXiv:0804.2044 [hep-ph]} \BibitemShut
  {NoStop}%
%%CITATION = ARXIV:0804.2044;%%
\bibitem [{\citenamefont {Hayakawa}\ \emph {et~al.}(2016)\citenamefont
  {Hayakawa}, \citenamefont {Blum}, \citenamefont {Christ}, \citenamefont
  {Izubuchi}, \citenamefont {Jin},\ and\ \citenamefont
  {Lehner}}]{Hayakawa:2015ntr}%
  \BibitemOpen
  \bibfield  {author} {\bibinfo {author} {\bibfnamefont {M.}~\bibnamefont
  {Hayakawa}}, \bibinfo {author} {\bibfnamefont {T.}~\bibnamefont {Blum}},
  \bibinfo {author} {\bibfnamefont {N.~H.}\ \bibnamefont {Christ}}, \bibinfo
  {author} {\bibfnamefont {T.}~\bibnamefont {Izubuchi}}, \bibinfo {author}
  {\bibfnamefont {L.~C.}\ \bibnamefont {Jin}}, \ and\ \bibinfo {author}
  {\bibfnamefont {C.}~\bibnamefont {Lehner}},\ }\bibfield  {booktitle} {\emph
  {\bibinfo {booktitle} {{Proceedings, 33rd International Symposium on Lattice
  Field Theory (Lattice 2015): Kobe, Japan, July 14-18, 2015}}},\ }\href@noop
  {} {\bibfield  {journal} {\bibinfo  {journal} {PoS}\ }\textbf {\bibinfo
  {volume} {LATTICE2015}},\ \bibinfo {pages} {104} (\bibinfo {year} {2016})},\
  \Eprint {http://arxiv.org/abs/1511.01493} {arXiv:1511.01493 [hep-lat]}
  \BibitemShut {NoStop}%
%%CITATION = ARXIV:1511.01493;%%
\bibitem [{\citenamefont {Blum}\ \emph
  {et~al.}(2016{\natexlab{d}})\citenamefont {Blum} \emph
  {et~al.}}]{Blum:2014tka}%
  \BibitemOpen
  \bibfield  {author} {\bibinfo {author} {\bibfnamefont {T.}~\bibnamefont
  {Blum}} \emph {et~al.} (\bibinfo {collaboration} {RBC, UKQCD}),\ }\href
  {\doibase 10.1103/PhysRevD.93.074505} {\bibfield  {journal} {\bibinfo
  {journal} {Phys. Rev.}\ }\textbf {\bibinfo {volume} {D93}},\ \bibinfo {pages}
  {074505} (\bibinfo {year} {2016}{\natexlab{d}})},\ \Eprint
  {http://arxiv.org/abs/1411.7017} {arXiv:1411.7017 [hep-lat]} \BibitemShut
  {NoStop}%
%%CITATION = ARXIV:1411.7017;%%
\bibitem [{\citenamefont {Blum}\ \emph {et~al.}(2013)\citenamefont {Blum},
  \citenamefont {Izubuchi},\ and\ \citenamefont {Shintani}}]{Blum:2012uh}%
  \BibitemOpen
  \bibfield  {author} {\bibinfo {author} {\bibfnamefont {T.}~\bibnamefont
  {Blum}}, \bibinfo {author} {\bibfnamefont {T.}~\bibnamefont {Izubuchi}}, \
  and\ \bibinfo {author} {\bibfnamefont {E.}~\bibnamefont {Shintani}},\ }\href
  {\doibase 10.1103/PhysRevD.88.094503} {\bibfield  {journal} {\bibinfo
  {journal} {Phys. Rev.}\ }\textbf {\bibinfo {volume} {D88}},\ \bibinfo {pages}
  {094503} (\bibinfo {year} {2013})},\ \Eprint {http://arxiv.org/abs/1208.4349}
  {arXiv:1208.4349 [hep-lat]} \BibitemShut {NoStop}%
%%CITATION = ARXIV:1208.4349;%%
\bibitem [{\citenamefont {Shintani}\ \emph {et~al.}(2015)\citenamefont
  {Shintani}, \citenamefont {Arthur}, \citenamefont {Blum}, \citenamefont
  {Izubuchi}, \citenamefont {Jung},\ and\ \citenamefont
  {Lehner}}]{Shintani:2014vja}%
  \BibitemOpen
  \bibfield  {author} {\bibinfo {author} {\bibfnamefont {E.}~\bibnamefont
  {Shintani}}, \bibinfo {author} {\bibfnamefont {R.}~\bibnamefont {Arthur}},
  \bibinfo {author} {\bibfnamefont {T.}~\bibnamefont {Blum}}, \bibinfo {author}
  {\bibfnamefont {T.}~\bibnamefont {Izubuchi}}, \bibinfo {author}
  {\bibfnamefont {C.}~\bibnamefont {Jung}}, \ and\ \bibinfo {author}
  {\bibfnamefont {C.}~\bibnamefont {Lehner}},\ }\href {\doibase
  10.1103/PhysRevD.91.114511} {\bibfield  {journal} {\bibinfo  {journal} {Phys.
  Rev.}\ }\textbf {\bibinfo {volume} {D91}},\ \bibinfo {pages} {114511}
  (\bibinfo {year} {2015})},\ \Eprint {http://arxiv.org/abs/1402.0244}
  {arXiv:1402.0244 [hep-lat]} \BibitemShut {NoStop}%
%%CITATION = ARXIV:1402.0244;%%
\bibitem [{\citenamefont {Mcglynn}(2016)}]{Mcglynn:2015uwh}%
  \BibitemOpen
  \bibfield  {author} {\bibinfo {author} {\bibfnamefont {G.}~\bibnamefont
  {Mcglynn}},\ }\bibfield  {booktitle} {\emph {\bibinfo {booktitle}
  {{Proceedings, 33rd International Symposium on Lattice Field Theory (Lattice
  2015): Kobe, Japan, July 14-18, 2015}}},\ }\href@noop {} {\bibfield
  {journal} {\bibinfo  {journal} {PoS}\ }\textbf {\bibinfo {volume}
  {LATTICE2015}},\ \bibinfo {pages} {019} (\bibinfo {year} {2016})}\BibitemShut
  {NoStop}%
%%CITATION = POSCI,LATTICE2015,019;%%
\bibitem [{\citenamefont {Yin}\ and\ \citenamefont
  {Mawhinney}(2011)}]{Yin:2011np}%
  \BibitemOpen
  \bibfield  {author} {\bibinfo {author} {\bibfnamefont {H.}~\bibnamefont
  {Yin}}\ and\ \bibinfo {author} {\bibfnamefont {R.~D.}\ \bibnamefont
  {Mawhinney}},\ }\bibfield  {booktitle} {\emph {\bibinfo {booktitle}
  {{Proceedings, 29th International Symposium on Lattice field theory (Lattice
  2011): Squaw Valley, Lake Tahoe, USA, July 10-16, 2011}}},\ }\href@noop {}
  {\bibfield  {journal} {\bibinfo  {journal} {PoS}\ }\textbf {\bibinfo {volume}
  {LATTICE2011}},\ \bibinfo {pages} {051} (\bibinfo {year} {2011})},\ \Eprint
  {http://arxiv.org/abs/1111.5059} {arXiv:1111.5059 [hep-lat]} \BibitemShut
  {NoStop}%
%%CITATION = ARXIV:1111.5059;%%
\bibitem [{\citenamefont {Bijnens}\ and\ \citenamefont
  {Relefors}(2016)}]{Bijnens:2016hgx}%
  \BibitemOpen
  \bibfield  {author} {\bibinfo {author} {\bibfnamefont {J.}~\bibnamefont
  {Bijnens}}\ and\ \bibinfo {author} {\bibfnamefont {J.}~\bibnamefont
  {Relefors}},\ }\href@noop {} {\  (\bibinfo {year} {2016})},\ \Eprint
  {http://arxiv.org/abs/1608.01454} {arXiv:1608.01454 [hep-ph]} \BibitemShut
  {NoStop}%
%%CITATION = ARXIV:1608.01454;%%
\bibitem [{\citenamefont {Jin}\ \emph {et~al.}(2016{\natexlab{a}})\citenamefont
  {Jin}, \citenamefont {Blum}, \citenamefont {Christ}, \citenamefont
  {Hayakawa}, \citenamefont {Izubuchi},\ and\ \citenamefont
  {Lehner}}]{Jin:2016xxx}%
  \BibitemOpen
  \bibfield  {author} {\bibinfo {author} {\bibfnamefont {L.}~\bibnamefont
  {Jin}}, \bibinfo {author} {\bibfnamefont {T.}~\bibnamefont {Blum}}, \bibinfo
  {author} {\bibfnamefont {N.}~\bibnamefont {Christ}}, \bibinfo {author}
  {\bibfnamefont {M.}~\bibnamefont {Hayakawa}}, \bibinfo {author}
  {\bibfnamefont {T.}~\bibnamefont {Izubuchi}}, \ and\ \bibinfo {author}
  {\bibfnamefont {C.}~\bibnamefont {Lehner}},\ }\bibfield  {booktitle} {\emph
  {\bibinfo {booktitle} {{Proceedings, 34rd International Symposium on Lattice
  Field Theory (Lattice 2016): Southampton, UK, July 24-30, 2016}}},\
  }\href@noop {} {\bibfield  {journal} {\bibinfo  {journal} {PoS}\ }\textbf
  {\bibinfo {volume} {LATTICE2016}},\ \bibinfo {pages} {???} (\bibinfo {year}
  {2016}{\natexlab{a}})},\ \bibinfo {note} {to appear}\BibitemShut {NoStop}%
\bibitem [{\citenamefont {Jin}\ \emph {et~al.}(2016{\natexlab{b}})\citenamefont
  {Jin}, \citenamefont {Blum}, \citenamefont {Christ}, \citenamefont
  {Hayakawa}, \citenamefont {Izubuchi},\ and\ \citenamefont
  {Lehner}}]{Jin:2015bty}%
  \BibitemOpen
  \bibfield  {author} {\bibinfo {author} {\bibfnamefont {L.}~\bibnamefont
  {Jin}}, \bibinfo {author} {\bibfnamefont {T.}~\bibnamefont {Blum}}, \bibinfo
  {author} {\bibfnamefont {N.}~\bibnamefont {Christ}}, \bibinfo {author}
  {\bibfnamefont {M.}~\bibnamefont {Hayakawa}}, \bibinfo {author}
  {\bibfnamefont {T.}~\bibnamefont {Izubuchi}}, \ and\ \bibinfo {author}
  {\bibfnamefont {C.}~\bibnamefont {Lehner}},\ }\bibfield  {booktitle} {\emph
  {\bibinfo {booktitle} {{Proceedings, 33rd International Symposium on Lattice
  Field Theory (Lattice 2015): Kobe, Japan, July 14-18, 2015}}},\ }\href@noop
  {} {\bibfield  {journal} {\bibinfo  {journal} {PoS}\ }\textbf {\bibinfo
  {volume} {LATTICE2015}},\ \bibinfo {pages} {103} (\bibinfo {year}
  {2016}{\natexlab{b}})},\ \Eprint {http://arxiv.org/abs/1511.05198}
  {arXiv:1511.05198 [hep-lat]} \BibitemShut {NoStop}%
%%CITATION = ARXIV:1511.05198;%%
\bibitem [{\citenamefont {Green}\ \emph {et~al.}(2016)\citenamefont {Green},
  \citenamefont {Asmussen}, \citenamefont {Gryniuk}, \citenamefont {von
  Hippel}, \citenamefont {Meyer}, \citenamefont {Nyffeler},\ and\ \citenamefont
  {Pascalutsa}}]{Green:2015mva}%
  \BibitemOpen
  \bibfield  {author} {\bibinfo {author} {\bibfnamefont {J.}~\bibnamefont
  {Green}}, \bibinfo {author} {\bibfnamefont {N.}~\bibnamefont {Asmussen}},
  \bibinfo {author} {\bibfnamefont {O.}~\bibnamefont {Gryniuk}}, \bibinfo
  {author} {\bibfnamefont {G.}~\bibnamefont {von Hippel}}, \bibinfo {author}
  {\bibfnamefont {H.~B.}\ \bibnamefont {Meyer}}, \bibinfo {author}
  {\bibfnamefont {A.}~\bibnamefont {Nyffeler}}, \ and\ \bibinfo {author}
  {\bibfnamefont {V.}~\bibnamefont {Pascalutsa}},\ }\bibfield  {booktitle}
  {\emph {\bibinfo {booktitle} {{Proceedings, 33rd International Symposium on
  Lattice Field Theory (Lattice 2015): Kobe, Japan, July 14-18, 2015}}},\
  }\href@noop {} {\bibfield  {journal} {\bibinfo  {journal} {PoS}\ }\textbf
  {\bibinfo {volume} {LATTICE2015}},\ \bibinfo {pages} {109} (\bibinfo {year}
  {2016})},\ \Eprint {http://arxiv.org/abs/1510.08384} {arXiv:1510.08384
  [hep-lat]} \BibitemShut {NoStop}%
%%CITATION = ARXIV:1510.08384;%%
\bibitem [{\citenamefont {Asmussen}\ \emph {et~al.}(2016)\citenamefont
  {Asmussen}, \citenamefont {Green}, \citenamefont {Meyer},\ and\ \citenamefont
  {Nyffeler}}]{Asmussen:2016lse}%
  \BibitemOpen
  \bibfield  {author} {\bibinfo {author} {\bibfnamefont {N.}~\bibnamefont
  {Asmussen}}, \bibinfo {author} {\bibfnamefont {J.}~\bibnamefont {Green}},
  \bibinfo {author} {\bibfnamefont {H.~B.}\ \bibnamefont {Meyer}}, \ and\
  \bibinfo {author} {\bibfnamefont {A.}~\bibnamefont {Nyffeler}},\ }\href@noop
  {} {\  (\bibinfo {year} {2016})},\ \Eprint {http://arxiv.org/abs/1609.08454}
  {arXiv:1609.08454 [hep-lat]} \BibitemShut {NoStop}%
%%CITATION = ARXIV:1609.08454;%%
\bibitem [{\citenamefont {Boyle}(2009)}]{Boyle:2009vp}%
  \BibitemOpen
  \bibfield  {author} {\bibinfo {author} {\bibfnamefont {P.~A.}\ \bibnamefont
  {Boyle}},\ }\href {\doibase 10.1016/j.cpc.2009.08.010} {\bibfield  {journal}
  {\bibinfo  {journal} {Comput. Phys. Commun.}\ }\textbf {\bibinfo {volume}
  {180}},\ \bibinfo {pages} {2739} (\bibinfo {year} {2009})}\BibitemShut
  {NoStop}%
%%CITATION = CPHCB,180,2739;%%
\bibitem [{\citenamefont {Jung}(2014)}]{Jung:2014ata}%
  \BibitemOpen
  \bibfield  {author} {\bibinfo {author} {\bibfnamefont {C.}~\bibnamefont
  {Jung}} (\bibinfo {collaboration} {RBC, UKQCD}),\ }\bibfield  {booktitle}
  {\emph {\bibinfo {booktitle} {{Proceedings, 31st International Symposium on
  Lattice Field Theory (Lattice 2013)}}},\ }\href@noop {} {\bibfield  {journal}
  {\bibinfo  {journal} {PoS}\ }\textbf {\bibinfo {volume} {LATTICE2013}},\
  \bibinfo {pages} {417} (\bibinfo {year} {2014})}\BibitemShut {NoStop}%
%%CITATION = POSCI,LATTICE2013,417;%%
\end{thebibliography}%

\end{document}